\documentclass[pre,aps,showpacs,twocolumn]{revtex4}

\usepackage{graphicx}
\usepackage{dcolumn}
\usepackage{bm}

\begin{document}


\title{Estimate for the fractal dimension of the Apollonian gasket in $d$
dimensions}

\author{R. S. Farr}
 \affiliation{Unilever R\&D, Olivier van Noortlaan 120, AT3133, Vlaardingen, The Netherlands and
 the London Institute for Mathematical Sciences, 22 S. Audley St., Mayfair, London, UK}
 \email{robert.farr@unilever.com}
\author{E. Griffiths}
 \noaffiliation
 \email{egriff70@gmail.com}

\date{\today}

\begin{abstract}
We adapt a recent theory for the random close packing of
polydisperse spheres in three dimensions [R. S. Farr and R. D. Groot, 
J. Chem. Phys. {\bf 131} 244104 (2009)] in order to predict the
Hausdorff dimension $d_{A}$ of the Apollonian gasket in dimensions 2 and above.
Our approximate results agree with published values in $2$ and $3$
dimensions to within $0.05\%$ and $0.6\%$ respectively, and we provide 
predictions for dimensions $4$ to $8$.
\end{abstract}

\pacs{05.45.Df, 61.43.Gt, 61.43.Hv}

\maketitle

\begin{figure}
\includegraphics[width=3in]{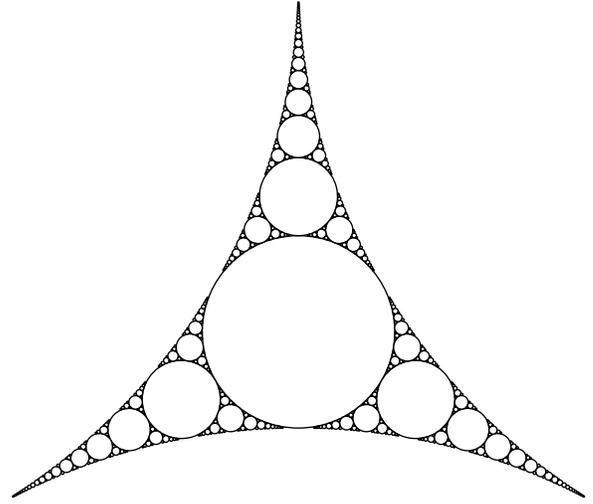}
\caption{\label{apol}
Apollonian packing of discs.
}
\end{figure}

Leibniz \cite{Leibniz} first suggested that a plane area can
be completely covered with discs, in an approximately self-similar manner,
through a construction which
involves starting with three equal touching discs, and then repeatedly
adding the largest possible disc which touches three neighbours, but
does not overlap with any disc already in the packing. The result is
illustrated in figure \ref{apol}.
According to Pappus of Alexandria, the problem
of finding such osculating circles was first studied by Apollonius of
Perga, in whose honour this `Apollonian packing' is named.
A similar construction can be envisaged for spheres (where each added
sphere touches four neighbours \cite{Borkovec}). In higher
dimensions, a construction based upon iterating the analogue
of Soddy's formula \cite{Soddy} or applying iterated 
inversions \cite{Mandelbrot} to hyperspheres will lead to 
overlaps \cite{Moraal}. Therefore in this paper we use the 
term `Apollonian packing'
to refer to an `osculatory packing' \cite{Borkovec}, which starts from $d+1$ equal,
touching hyperspheres at the vertices of a regular $d-$simplex, and where 
repeatedly, the largest possible
$d$-dimensional hypersphere is added to the
existing packing that does not overlap any already present. The added
hypersphere touches $d+1$ others at this stage (although this fact is
not needed for the subsequent argument).

Apollonian networks \cite{Doye}, which are graphs derived 
from Apollonian packings, have been suggested as models for 
real-world networks, such as social networks and
hierarchical road systems \cite{Andrade}. In these contexts, 
networks based on Apollonian packings with any dimension,
including $d>3$ may be of practical relevance \cite{Zhang}.

In lower dimensions ($d=2,3$), the physical significance of 
Apollonian packings is that they can be used as
idealized models of high density granular materials, for example
in high strength concrete \cite{Hermann}. Furthermore, related 
constructions, such
as space-filling bearings in two \cite{MannaVicsek} and three 
\cite{Baram} dimensions, and random space filling 
bearings \cite{Baram2} have
been proposed as simplified models for the geometry of turbulence
\cite{Bachelor} or the broken material near a geological fault.
Random Apollonian packings of shapes other than hyperspheres
have also been studied \cite{Delaney}. In all these cases,
the method of adding spheres is modified so that it is no
longer the largest possible non-overlapping sphere which is added
at each stage; for example in the case
of bearings an additional constraint is needed to ensure a bichromatic 
colouring \cite{Baram}. These
modifications all have the effect of reducing the rapidity with 
which the packing approaches a volume fraction of unity as 
spheres are added and also alter the fractal dimension of
their residual sets. In recent work on random bearings
\cite{Lind} the fractal dimension can even be varied continuously
by choice of parameters.

The residual set or `Apollonian 
gasket' of such structures is of practical relevance, since its surface 
area and volume (for the 3d case) are related to solvent adsorption and 
permeability to  flow through the packing. These 
geometrical properties of the residual set
are finite, provided the packing contains only spheres larger than a 
certain cutoff diameter. However, in the limit where spheres of arbitrarily
small size are included, the residual set is fractal in nature, and 
its Hausdorff dimension captures the essential
geometric information \cite{Mandelbrot}.

Recent high-precision calculations have shown that in 2d, the dimension
of the Apollonian gasket is $d_{A,2}\approx 1.30568$ \cite{Manna}, while
in 3d, it is $d_{A,3}\approx 2.4739465$ \cite{Borkovec}.

Since the Apollonian packing is a special kind of sphere packing 
in three dimensions, then it is interesting to investigate whether 
the recent approximate
theory for the volume fraction of random close packings of polydisperse spheres,
presented in Ref. \cite{Farr} may shed some light on this problem also.
The hope is that the essential geometric features of sphere packings captured in 
Ref. \cite{Farr} might apply to non-random cases also.

In the theory of Ref. \cite{Farr}, we start with a known distribution of 
sphere diameters $P_{3d}(D)$, where $P_{3d}(D){\rm d}D$ is the 
number fraction of the spheres with diameters in the range $(D,D+{\rm d}D)$,
and we ask what is the maximum random packing density which can be obtained?

The procedure consists of several stages: First, $P_{3d}(D)$ is converted 
into a number distribution of
one dimensional rods $P_{1d}(L)$, by imagining a random non-overlapping 
(but not necessarily close 
packed) distribution of spheres, passing a straight line through this
distribution, and counting each portion of the line within a sphere to be a 
rod. The resulting distribution of rod lengths is given \cite{Farr} by
\begin{equation}\label{eq1}
P_{1d}(L)=2L\frac{\int_{L}^{\infty}P_{3d}(D){\rm d}D}{
\int_{0}^{\infty}P_{3d}(D)D^{2}{\rm d}D}.
\end{equation}

In order to simulate packing, we then imagine that this collection of rods 
interacts on a line through a hard pair potential which acts
between each pair of rods $L_{i}$ and $L_{j}$ through
\begin{equation}\label{eq2}
V(h)=\left\{
\begin{array}{lll}
\infty & {\rm if} & h < \min(fL_{i},fL_{j}) \\
0 & {\rm if} & h \ge \min(fL_{i},fL_{j})
\end{array}\right.
\end{equation}
In Eq. (\ref{eq2}), $h$ is the closest approach of the two ends of the rods,
$f>0$ is a free parameter in the theory (which we explain later), and the
potential is able to reach through smaller rods which may be in the 
gap between the two rods under consideration.

Finally, we search over all orderings of the rods, and find the ordering
which occupies the maximum length fraction on the line. This search
can be accomplished by a simple greedy algorithm \cite{Farr}, and the
final length fraction occupied by the rods is our estimate for the
maximum random packing fraction of the spheres in 3d.

As described, the theory depends on a free parameter $f$, which in
Ref. \cite{Farr} is fixed by ensuring that the predicted close
packing density for monodisperse spheres matches the known random
close packing density $\phi_{\rm RCP}\approx 0.6435$. With this
calibration, $f\approx 0.7654$ and the theory can be applied to 
arbitrary sphere size distributions.

For the further development of this paper, we require the generalization of 
this model to other dimensions. Therefore consider a polydisperse
collection of $d-$dimensional hyperspheres, where $P_{dd}(D){\rm d}D$ 
is the number fraction of hyperspheres with diameters in the 
range $(D,D+{\rm d}D)$,
and $d\ge 2$. If we consider passing a straight line at random through a 
single hypersphere of diameter $D$, then we will generate a collection
of rods with a normalized length distribution given by
\begin{equation}\label{mono}
\hat{p}_{1d}(L;D)=(d-1)L D^{1-d}(D^2 - L^2)^{(d-3)/2}\theta(D-L),
\end{equation}
where $\theta(x)$ is the Heaviside step function.

The distribution of rod lengths generated from passing a line through
a random distribution of $d$ dimensional hyperspheres, will therefore
be given by a convolution with $P_{dd}$, but also taking into
account that the collision cross section for the line with a hypersphere
of diameter $D$ scales as $D^{d-1}$. The result is
\begin{equation}
P_{1d}(x)\propto \int_{D=L}^{\infty} D^{d-1}P_{dd}(D) \hat{p}_{1d}(L;D){\rm d}D,
\end{equation}
which with the correct normalization (obtained by reversing the order of
integration over $D$ and $L$), gives the appropriate generalization
of Eq. (\ref{eq1}), namely
\begin{equation}\label{rods}
P_{1d}(L)=(d-1)L\frac{\int_{L}^{\infty}(D^2 - L^2 )^{(d-3)/2}
P_{dd}(D){\rm d}D}{\int_{0}^{\infty}
D^{d-1}P_{dd}(D){\rm d}D}.
\end{equation}

\begin{figure}
\includegraphics[width=3in]{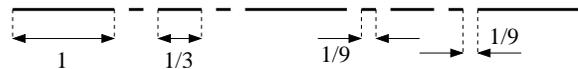}
\caption{\label{ap_rods}
Analogue of Apollonian packing for rods on a line, using the
potential of Eq. (\ref{eq2}) in the text with $f=1$, and two iterations
of fitting small rods into the gaps between large rods.
}
\end{figure}

Now, consider the analogue of the Apollonian packing for rods on a line
subject to the potential of Eq. (\ref{eq2}). This consists of starting with 
a set of equal large rods, and then placing the largest possible rods
into the gaps between them, which do not require the large rods to move.
This process is then repeated iteratively, as in figure \ref{ap_rods}.

In one unit cell of this structure, there is one rod of the longest length
(which we take as unity), which leaves a gap of size $f$ to be filled
by the smaller rods. In choosing and placing these smaller rods, we need
$2^{0}$ rods of length $f(1+2f)^{-1}$, then in the remaining gaps, which
are of length $f^{2}(1+2f)^{-1}$, we place $2^{1}$ rods of length
$f^{2}(1+2f)^{-2}$. Repeating this process, we have at iteration number
$j$, $2^{j}$ rods of length $f^{j+1}(1+2f)^{-(j+1)}$.

This implies that asymptotically, as $L\rightarrow 0$, the total number
of rods of size greater than $L$ behaves as
\begin{equation}
\int_{L}^{\infty}P_{1d}(L'){\rm d}L'\propto 1+\sum_{i=0}^{j}2^{i},
\end{equation}
where $L=f^{j+1}(1+2f)^{-(j+1)}$. Therefore
\begin{equation}\label{x}
\int_{L}^{\infty}P_{1d}(L'){\rm d}L'\propto L^{x}\ \ {\rm where}\ \ 
x=\frac{\ln 2}{\ln\left(\frac{f}{1+2f}\right)}.
\end{equation}
Now, the distribution $P_{1d}(L)$ in Eq. (\ref{x}) has a corresponding 
distribution $P_{dd}(D)$ of hyperspheres, which from Eq. (\ref{rods})
is given asymptotically in the limit $D\rightarrow 0$ by
\begin{equation}\label{y}
\int_{D}^{\infty}P_{dd}(D'){\rm d}D'\propto D^{y}\ \ {\rm where}\ \ 
y=x-d+1.
\end{equation}

We now link these results back to the dimension $d_{A}$ of the
Apollonian gasket in $d$ dimensions in the following manner: According to Refs. 
\cite{Borkovec,Boyd1,Boyd2}, the cardinality of the set of spheres
in an Apollonian packing, with curvature not exceeding $\kappa$
is given by
\begin{equation}\label{boyd}
N(\kappa)\propto \kappa^{d_{A}}.
\end{equation}

Combining Eqs. (\ref{x}), (\ref{y}) and (\ref{boyd}), we therefore obtain 
our estimate for the Hausdorff dimension $d_{A}$ of the residual set of the 
packing: 
\begin{equation}\label{dA_approx}
d_{A}\approx -y = d-1-\frac{\ln 2}{\ln\left(\frac{f}{1+2f}\right)},
\end{equation}
where $f$ is the appropriate value for each dimension $d$ of space.

To complete the calculation, we need a value for the free 
parameter $f$ in the theory.
This will be done (as in the analysis of random close packing \cite{Farr}) 
by calibrating the theory for the monodisperse case.
To obtain the rod distribution corresponding to monodisperse
hyperspheres, we use 
a collection of $n=50\ 000$ rods sampled uniformly from the 
distribution of Eq. (\ref{mono}). To do this, we take equal points on the
inverse function of the integral of Eq. (\ref{mono}), so that our rod
lengths are given by
\begin{equation}
L_{i}=D\left[ 1- \left(\frac{i}{n}-\frac{1}{2n}\right)^{2/(d-1)}\right]^{1/2},
\end{equation}
where $i=1\ldots n$.

\begin{figure}
\includegraphics[width=3in]{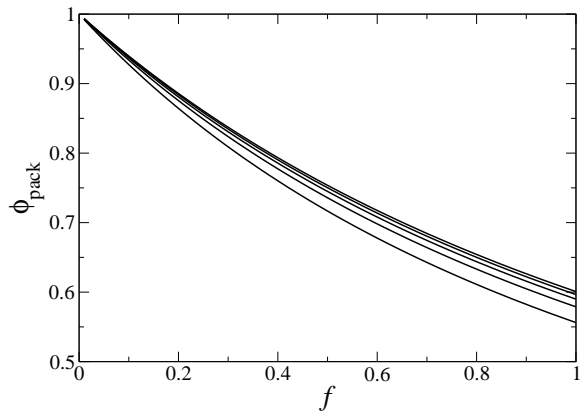}
\caption{\label{fcalib}
Plot of the maximum packing fraction $\phi_{\rm pack}$ of $d$ dimensional 
hyperspheres, as a function of the parameter $f$. The curves from bottom
to top correspond to $d=2$, $3$, $4$, $5$ and $6$.
}
\end{figure}

Applying the greedy one-dimensional packing algorithm described 
in Ref. \cite{Farr}, we calculate numerically the packing density of monodisperse
hyperspheres as a function of the parameter $f$, for each dimension
$d$ of space. The resulting curves for $d=2$ to $6$ are shown in
figure \ref{fcalib}.

In order to apply the packing theory, we need the value of $f$ for
each dimension $d$. When applying the theory to random close packing in 3d, the
calibration used was to ensure that the prediction for random close packing
of monodisperse spheres was correct \cite{Farr}.

\begin{figure}
\includegraphics[width=3in]{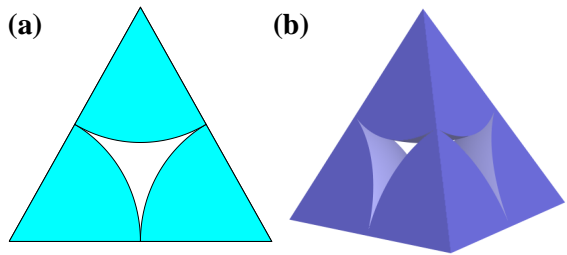}
\caption{\label{trifig}
(a) The maximum packing fraction $\phi_{\rm simp}$ for $d=2$ is the ratio 
of the shaded area to that inside the complete 
regular $2$-simplex (equilateral triangle). (b) The same construction for $d=3$.
}
\end{figure}

For the Apollonian packing each hypersphere is added in such a way
as to optimally fill the remaining available space, and
so the local geometry of packing will always be as efficient as possible.
In order to capture this property, we choose $f$ to give the maximum 
possible local packing fraction of equal hyperspheres. This corresponds
to placing equal osculating hyperspheres at the vertices of a regular
$d$-simplex, and calculating the volume fraction occupied inside
the simplex. We refer to this packing fraction as $\phi_{\rm simp}$, and
it is illustrated for the cases $d=2$ and $3$ in figure \ref{trifig}.
An alternative argument for this choice, is that 
we are calibrating $f$ by using the true packing fraction of the first few
hyperspheres in the Apollonian packing. In general, these will be of 
different sizes, but by taking the first $d+1$, we again only
need to consider equal spheres at the vertices of a simplex.

In 2d, this construction gives the same area fraction as a hexagonal packing, so
$\phi_{\rm simp}=\pi/(2\sqrt{3})\approx 0.9069$. In three dimensions, we find 
$\phi_{\rm simp}=3\sqrt{2}\left[ \cos^{-1}(1/3)-\pi/3\right]\approx 0.7796$, which is higher 
than can be achieved
for any global packing of spheres in 3d (this limit is 
$\phi_{\rm fcc}=\pi/\sqrt{18}\approx 0.74$, achieved for the face centred cubic or hexagonal close
packed arrangement \cite{Hales}).

For higher dimensions, we calculate the simplex packing fraction using
a Monte-Carlo integration, noting that if one vertex of a regular
simplex lies at the origin of $d$-dimensional Cartesian coordinates,
then the other vertices can be chosen at the 
positions $\{ {\rm\bf s}^{j}\}$ where
\begin{equation}
s^{j}_{i}=\left( 2d+4 +4\sqrt{1+d}\right)^{-1/2}\left[ \left( 1+\sqrt{1+d}
\right)\delta_{i,j} +1 \right].
\end{equation}
A point ${\rm\bf p}$ chosen randomly (and uniformly) in $(0,1)^{d}$ can be
expanded as ${\rm\bf p}=\sum_{j}q_{j}{\rm\bf s}^{j}$, where
$q_{j}=\sum_{i}t_{j}^{i}p_{i}$ and
\begin{equation}
t_{j}^{i}=\frac{( 2d+4 +4\sqrt{1+d})^{1/2}}{
1+\sqrt{1+d}}
\left[
\delta_{i,j}-\frac{1}{1+d+\sqrt{1+d}}\right].
\end{equation}
The point ${\rm\bf p}$ lies within the simplex if all the $q_{j}$'s
are positive, and their sum does not exceed unity. We denote the volume of 
this simplex by $V_{\rm simp}$, which can thus be obtained by a Monte-Carlo
integration, or from the
analytic expression 
$V_{\rm simp}=2^{-d/2}\sqrt{(1+d)}/d!$.
Furthermore, consider the volume $V_{\rm sph}$ of that portion of a 
unit radius
hypersphere lying within a
large regular $d$-simplex, when the hypersphere has its centre at one of the
vertices of the simplex. The point ${\rm\bf p}$ lies within this volume
if all the $q_{j}$'s
are positive, and $\sum_{j}(q_{j})^{2}< 1$. Again, this allows us to calculate
$V_{\rm sph}$ stochastically.

From these two quantities, the maximum packing
fraction in a simplex is given by
\begin{equation}
\phi_{\rm simp}=(d+1)V_{\rm sph}/\left( 2^{d}V_{\rm simp}\right),
\end{equation}
which is shown in table \ref{gasket}, alongside the predicted 
values of $d_{A}$ [from Eq. (\ref{dA_approx})] and the
actual values (where known).

\begin{table}
\caption{\label{gasket}
Predictions for the Hausdorff dimension of the Apollonian gasket
in $d$ dimensions. The close packing density of spheres with centres
at the vertices of a regular $d$-simplex is $\phi_{\rm simp}$. The
corresponding value of $f$ from the packing theory is shown, along with the
predicted Hausdorff dimension $d_{A}^{\rm pred}$, and the actual
Hausdorff dimension $d_{A}^{\rm act}$ if known \cite{Manna,Borkovec}.
}
\begin{tabular}{cllll}
\hline
\hspace{1em} $d$ \hspace{1em}  & $\phi_{\rm simp}$\hspace{1em} & $f$\hspace{1em} & $d_{A}^{\rm pred}$\hspace{1em} & $d_{A}^{\rm act}$\hspace{1em} \\
\hline
$2$ & $0.906900$ \ \ & $0.131025$ \ \ & $1.3060$ \ \ & $1.3057$ \\
$3$ & $0.779636$ & $0.394834$ & $2.4586$ & $2.4739$ \\
$4$ & $0.6478$ & $0.7864$ & $3.5848$ & \ \ \ - \\
$5$ & $0.5257$ & $1.325$ & $4.6840$ & \ \ \ - \\
$6$ & $0.4195$ & $2.047$ & $5.7603$ & \ \ \ - \\
$7$ & $0.330$ & $3.02$ & $6.8189$ & \ \ \ - \\
$8$ & $0.255$ & $4.35$ & $7.864$ & \ \ \ - \\
\hline
\end{tabular}
\end{table}

From table \ref{gasket}, we see that the predictions from this model
in $2$ and $3$
dimensions agree with the known values to within $0.05\%$ and $0.6\%$ 
respectively, and predictions for higher values of $d$ may be readily
obtained.

In conclusion, the packing theory of Ref. \cite{Farr}, which was designed to
abstract the essential geometric features of random close packing,
also appears to contain enough information to predict important
features of the hierarchical Apollonian packing. 
The extension of these arguments to more general Apollonian-type
packings (such as space filling bearings \cite{Baram} or random Apollonian
packings \cite{Delaney}) will require further work, because the objects
inserted into the packing are no longer maximal, which implies that both 
Eq. (\ref{x}) and the calibration of $f$ will need to be modified.
Nevertheless, we hope that
further study of this or related theories will lead to more insights and
further analytical results on both packings and granular materials.

\acknowledgments
Fig. \ref{apol} was supplied by 
the user `Time3000' on the `Wikimedia Commons' webproject.

\end{document}